\documentclass[%
 reprint,
 amsmath,amssymb,
 aps,
prb%,
,superscriptaddress
]{revtex4-1}

\usepackage{subfigure}
\usepackage{color}
\usepackage{graphicx}% Include figure files
\usepackage{tabularx}
\usepackage{dcolumn}% Align table columns on decimal point
\usepackage{bm}% bold math
\begin{document}

\preprint{APS/123-QED}

\title{Microscopic derivation of magnon spin current in a topological insulator/ferromagnet heterostructure}% Force line breaks with \\

%\author{Nobuyuki Okuma}
%\email{okuma@hosi.phys.s.u-tokyo.ac.jp}
 %\altaffiliation{Department of Physics, University of Tokyo, Hongo 7-3-1, 113-0033, Japan}%Lines break automatically or can be forced with \\
%\author{Masao Ogata}%
%\affiliation{%
 %Department of Physics, University of Tokyo, Hongo 7-3-1, Tokyo 113-0033, Japan
%\author{Kentaro Nomura}
 %\altaffiliation{Department of Physics, University of Tokyo, Hongo 7-3-1, 113-0033, Japan}%Lines break automatically or can be forced with \\
%\author{Masao Ogata}%
%\affiliation{%
 %Institute for Materials Research, Tohoku University, Sendai 980-8577, Japan

\author{Nobuyuki Okuma}\email{okuma@hosi.phys.s.u-tokyo.ac.jp}
\affiliation{Department of Physics, University of Tokyo, Hongo 7-3-1, 113-0033, Japan}
\author{Kentaro Nomura}%\email{}
\affiliation{Institute for Materials Research, Tohoku University, Sendai 980-8577, Japan}

\if0%
\author{Charlie Author}
 \homepage{http://www.Second.institution.edu/~Charlie.Author}
\affiliation{
 Second institution and/or address\\
 This line break forced% with \\
}%
\affiliation{
 Third institution, the second for Charlie Author
}%
\author{Delta Author}
\affiliation{%
 Authors' institution and/or address\\
 This line break forced with \textbackslash\textbackslash
}%

\collaboration{CLEO Collaboration}%\noaffiliation
\fi%
\date{\today}% It is always \today, today,

\begin{abstract} 
We investigate a spin-electricity conversion effect in a topological insulator/ferromagnet heterostructure.
In the spin-momentum-locked surface state, an electric current generates nonequilibrium spin accumulation, which causes a spin-orbit torque that acts on the ferromagnet.
When spins in the ferromagnet are completely parallel to the accumulated spin, this spin-orbit torque is zero.
In the presence of spin excitations, however, a coupling between magnons and electrons enables us to obtain a nonvanishing torque.
In this paper, we consider a model of the heterostructure in which a three-dimensional magnon gas is coupled with a two-dimensional massless Dirac electron system at the interface.
We calculate the torque induced by an electric field, which can be interpreted as a magnon spin current, up to the lowest order of the electron-magnon interaction.
We derive the expressions for high and low temperatures and estimate the order of magnitude of the induced spin current for realistic materials at room temperature.
\begin{description}
\item[PACS numbers]
72.25.Pn, 73.20.-r, 75.76.+j
\end{description}

\end{abstract}

\pacs{}% PACS, the Physics and Astronomy
\maketitle

%\tableofcontents
\section{INTRODUCTION}
The surface of a three-dimensional topological insulator (TI) is described by gapless Dirac electrons that have their spin locked at a right angle to their momentum \cite{hasan,xlq}.
Owing to this property, known as spin-momentum locking, an electric current in the surface is spin polarized and 
generates nonequilibrium spin accumulation $\langle\bm{s}\rangle_{neq}$ whose direction is perpendicular to the electric current and parallel to the surface
(the Rashba-Edelstein effect \cite{edelstein,inoue,kato,silov}).

Recently, couplings between the spin-momentum-locked surface state and magnetism have been studied, both theoretically \cite{nomura,sakai,taguchi,yokoyama,mahfouzi2} and experimentally 
\cite{fan,mellnik,ywang,shiomi,hwang,yasuda,kondou,dankert,tian,jamali,jiang}.
%Recently, couplings between the spin-momentum-locked surface state and magnetism have attracted much attention in spintronics \cite{nomura,yasuda,fan,mellnik,shiomi,hwang,kondou,dankert, ywang, tian,jiang,sakai,fischer,ndiaye,taguchi,yokoyama}.
In magnetically doped TIs under an electric field, spin accumulation on the surface state causes a spin-orbit torque \cite{chernyshov,miron} $\mathcal{T}\propto\langle\bm{s}\rangle_{neq}\times \hat{\bm{M}}$, where $\hat{\bm{M}}$ is the normalized magnetization vector of the magnetic dopants \cite{fan,mellnik,ywang}.
The anomalous Hall effect measurements in a (Bi$_{x}$Sb$_{1-x}$)$_2$Te$_3$/(Cr$_y$Bi$_{z}$Sb$_{1-y-z}$)$_2$Te$_3$ bilayer film have shown the existence of
the giant spin-orbit torque at the interface \cite{fan}.
In TI/ferromagnet (FM) heterostructures, a spin current injected by spin pumping is converted to an electric current via the inverse Rashba-Edelstein effect.
This phenomenon has been observed for a metallic FM permalloy \cite{shiomi} and for a magnetic insulator yttrium iron garnet (YIG) \cite{hwang}.

In this paper, we consider a TI/FM heterostructure in the presence of an electric field applied perpendicularly to the in-plane magnetization [Fig. \ref{fig1}(a)].
Although the conventional spin-orbit torque is zero, a coupling between electrons and low-energy spin excitations, known as magnons, enables us to obtain a nonvanishing torque.
We calculate microscopically this type of torque, which can be interpreted as a magnon spin current in the FM.
%Roughly speaking, this phenomenon is the inverse effect of the spin-pumping in the TI/FM heterostructure \cite{shiomi,hwang}.

This paper is organized as follows.
In Sec. \ref{model}, we define a model of the TI/FM heterostructure.
In our model, Dirac electrons in the two-dimensional surface state and magnons in the three-dimensional FM are coupled through the $s\mathchar`-d$ interaction at the interface.
In Sec. \ref{formulation}, we define the spin current operator at the interface and outline the calculation of the spin current. 
Based on the Kubo formalism, we evaluate the lowest-order contributions of the electron-magnon interaction.
In Sec. \ref{vertex}, we discuss the impurity vertex corrections. We conclude that the corrections to the electron-magnon vertexes are not important for large chemical potentials.
In Sec. \ref{explicit}, we derive the expressions of the spin current for high- and low- temperature limits.
In the derivation, we assume that the chemical potential is much larger than the other energy scales: temperature and magnon energies.
In Sec. \ref{dissum}, we discuss a quantitative estimate for realistic materials and summarize our work.

\section{MODEL\label{model}}
In this section, we describe a low-energy effective model of the TI/FM heterostructure [Fig. \ref{fig1}(a) and \ref{fig1}(b)].
In this paper, we set $\hbar=k_B=1$.

\subsection{Surface state}
A minimal Hamiltonian for the topological surface state is given by
\begin{align}
H_e&=\int\frac{d^2k}{(2\pi)^2}\psi_{\bm{k}}^{\dagger}\hat{\mathcal{H}}_e(\bm{k})\psi_{\bm{k}},\notag\\
\hat{\mathcal{H}}_e(\bm{k})&=-vk_x\hat{\sigma}_y+vk_y\hat{\sigma}_x-\mu \hat{1}\notag\\
&=\sum_{\alpha=\pm}\xi^{\alpha}_{\bm{k}}|\bm{k},\alpha\rangle\langle \bm{k},\alpha|,
\end{align}
where ($\psi,\psi^\dagger$) are the two-component spinors of the surface-state electrons, $\bm{k}=(k_x,k_y)$ is the electron momentum, $v$ is the Fermi velocity, $\mu>0$ is the chemical potential, and $\hat{\sigma}_i$ are the Pauli matrices in spin space.
In the second line, we define the projection operators $|\bm{k},\pm\rangle\langle \bm{k},\pm|=[\hat{1}\pm\bm{d}(\bm{k})\cdot\hat{\bm{\sigma}}]/2$
for the upper and lower bands with energies $\xi^{\pm}_{\bm{k}}=\pm v|\bm{k}|-\mu$, 
where $\bm{d}(\bm{k})=(\sin \theta_{\bm{k}},-\cos\theta_{\bm{k}},0)$, and $\theta_{\bm{k}}$ is the polar angle of the momentum $\bm{k}$.

In the following, we assume that the surface state is disordered by nonmagnetic impurities.
The thermal Green's function of electrons is given by
\begin{align}
\hat{\mathcal{G}}_{\bm{k}}(i\omega_n)&=\frac{1}{i\omega_n-\hat{\mathcal{H}}_e(\bm{k})-\hat{\Sigma}_{imp}(i\omega_n,\bm{k})}\notag\\
&\simeq\sum_{\alpha=\pm}|\bm{k},\alpha\rangle\langle \bm{k},\alpha|g_{\bm{k},\alpha}(i\omega_n),\label{electrongreen}
\end{align}
where $\omega_n=(2n+1)\pi T$, $T$ is the temperature, $\hat{\Sigma}_{imp}$ is the impurity self-energy, and $g_{\bm{k},\alpha}(i\omega_n)=[i\omega_n-\xi^{\alpha}_{\bm{k}}+\mathrm{sgn}(\omega_n)i/2\tau]^{-1}$.
In the second line, we use the relaxation time approximation and introduce the impurity relaxation time $\tau$. 
\subsection{Magnon gas}
\begin{figure}[t]
\begin{center}
　　　\includegraphics[width=8cm,angle=0,clip]{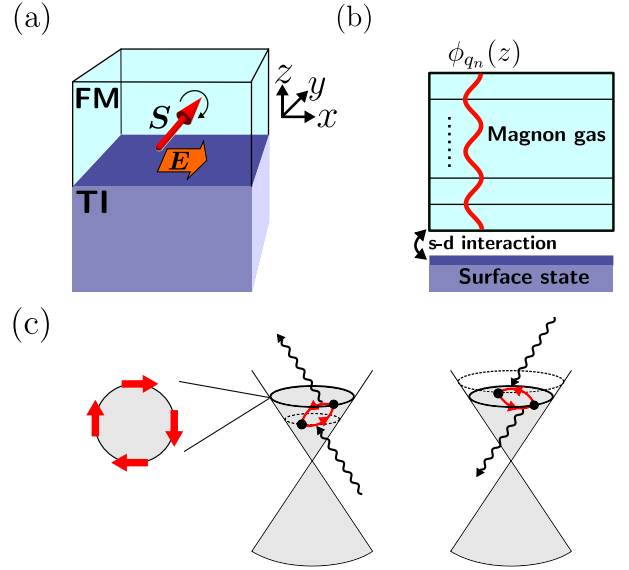}
　　　\caption{(a) A ferromagnet (FM) deposited on the surface of a topological insulator (TI). An applied electric field is perpendicular to spins in the FM. (b) Schematic picture of the model described in Sec. \ref{model}. 
The TI is described by a two-dimensional Dirac electron system on the surface state. 
The FM is described by magnons in three-dimensional space whose wave function in the $z$ direction, $\phi_{q_n}(z)$, obeys the Neumann boundary condition Eq. ($\ref{Neumann}$). The coupling between the TI and the FM is included through the $s$-$d$ interaction. (c) Electron-magnon-scattering processes on the spin-momentum-locked Fermi surface.}
　　　\label{fig1}
\end{center}
\end{figure}

For simplicity, we assume that the FM is described by an isotropic Heisenberg model.
We consider the case where spins in the FM are parallel to the $y$ direction, which is perpendicular to the electric field $\bm{E}=(E_x,0,0)$ [Fig. \ref{fig1}(a)]. 
The low-energy spin excitations of the FM are described by the magnon operators $(a, a^{\dagger})$, which are introduced by the spin-wave approximation: $S^y= S_0-a^{\dagger}a$, $S^z+iS^x\simeq\sqrt{2S_0}a$, and $S^z-iS^x\simeq\sqrt{2S_0}a^\dagger$, where $S^i$ and $S_0$ are the spin density operators and the magnitude of the spin density in the FM, respectively. 
In the following, we regard the FM as a three-dimensional magnon gas with a quadratic dispersion.
Using magnon operators, we obtain a low-energy effective Hamiltonian for a three-dimensional isotropic FM:
\begin{align}
H_{m}=\sum_{q_n}\int \frac{d^2q}{(2\pi)^2}\omega_{\bm{q},q_n}a^\dagger_{\bm{q},q_n}a_{\bm{q},q_n},
\end{align}
where $\bm{q}=(q_x,q_y)$ is the two-dimensional momentum, $q_n=n\pi/La$ ($n=0,1,\dots,L-1$) is the $z$ direction momentum, and $\omega_{\bm{q},q_n}=D(|\bm{q}|^2+q_n^2)$ is the magnon dispersion with the stiffness $D$. We assume that the system has $L$ sites with the lattice constant $a$ in the $z$-direction [Fig. \ref{fig1}(b)].
We also assume that the magnon wave function in the $z$-direction is given by $\phi_{q_n}(z)=\sqrt{2/L}\cos q_nz$, which obeys the Neumann boundary condition \cite{takei}:
\begin{align}
\partial_z\phi_{q_n}(z)|_{z=0}=\partial_z\phi_{q_n}(z)|_{z=La}=0.\label{Neumann}
\end{align}
Note that this boundary condition is approximately valid in the case where the interaction between electrons and magnons at the interface is small.
Using the above wave function, we obtain
\begin{align}
a^{(\dagger)}_{\bm{q}}(z)&=\sum_{q_n}\phi_{q_n}(z)a^{(\dagger)}_{\bm{q},q_n},\notag\\
S^{i}_{\bm{q}}(z)&=\sum_{q_n}\phi_{q_n}(z)S^{i}_{\bm{q},q_n},\label{expansion}
\end{align}
where $i=x,z$.
Assuming that the dissipation of the magnon gas is negligible, the thermal Green's function of magnons is given by 
\begin{align}
\mathcal{D}_{\bm{q},q_n}(i\omega_m)=\frac{1}{i\omega_m-\omega_{\bm{q},q_n}},\label{magnongreen}
\end{align}  
where $\omega_m=2\pi mT$.

\subsection{Electron-magnon interaction}

To include the interaction between the TI and the FM, 
we start with the $s$-$d$ Hamiltonian:
\begin{align}
 H_{sd}=-\frac{J_{sd}a^2}{2}\int dxdy\psi^\dagger(x,y)\hat{\bm{\sigma}}\psi(x,y)\cdot\bm{S}(x,y,z=0),
\end{align}
where $J_{sd}$ is the $s\mathchar`-d$ exchange coupling. In this Hamiltonian, the effect of the $y$ direction coupling is nothing other than the constant electron momentum shift in the $k_x$ direction, which does not affect transport. The remaining part can be rewritten as the following electron-magnon interaction:
\begin{align}
H_{sd}&=-\frac{J_{sd}a^2}{2}\sum_{i=x,z}\int\frac{d^2kd^2k'}{(2\pi)^2(2\pi)^2}  \psi^\dagger_{\bm{k}}\hat{\sigma}_i\psi_{\bm{k'}}S^i_{\bm{k'}-\bm{k}}(z=0)\notag\\
&=-\frac{J_{sd}a^2}{2}\sqrt{\frac{2}{L}}\sum_{\substack{i=x,z,\\q_n}}\int\frac{d^2kd^2k'}{(2\pi)^2(2\pi)^2} \psi^\dagger_{\bm{k}}\hat{\sigma}_i\psi_{\bm{k'}}S^i_{\bm{k'}-\bm{k},q_n},
\end{align}
where we use Eq. ($\ref{expansion}$).

\section{Formulation\label{formulation}}
\begin{figure}[]
\begin{center}
　　　\includegraphics[width=8cm,angle=0,clip]{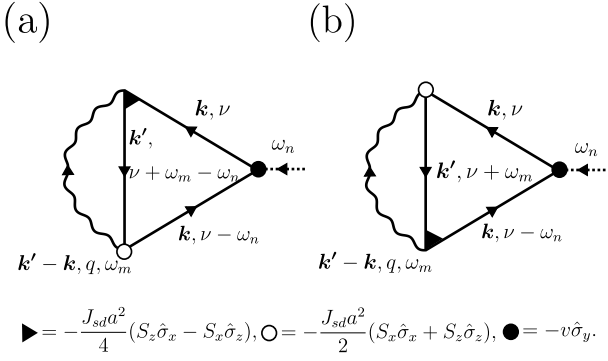}
　　　\caption{The diagrams contributing to $K^y(i\omega_n)$ at $J_{sd}^2$: (a) $K^{y(a)}(i\omega_n)$ and (b) $K^{y(b)}(i\omega_n)$.
The solid and wavy lines denote the electron green function $\mathcal{G}$ and spin-spin correlation function $\chi$, respectively.
The dotted line describes the external electric field.}
　　　\label{fig2}
\end{center}
\end{figure}

In this section, we outline the calculation of the spin current generated by the electron-magnon scattering.
In the following, we perform the perturbation calculation with respect to $J_{sd}$ by assuming $J_{sd}\ll\mu$.
The validity of this assumption is discussed in Sec. \ref{dissum}.
At the interface, the spin current is equivalent to the torque induced by the electron-magnon interaction, as discussed in the Introduction.
Thus, the spin current operator at the interface is given by \cite{takahashi}
\begin{align}
j^{S^y}_z&=\frac{1}{i}\left[\frac{S^y_{tot}}{V},H_{sd}\right]=-\frac{1}{i}\left[\frac{s^y_{tot}}{V},H_{sd}\right]\notag\\
&=\epsilon_{yjk}\frac{J_{sd}a^2}{4V}\sqrt{\frac{2}{L}}\sum_{q_n}\int\frac{d^2kd^2k'}{(2\pi)^2(2\pi)^2} \psi^\dagger_{\bm{k}}\hat{\sigma}_k\psi_{\bm{k'}}S^j_{\bm{k'}-\bm{k},q_n},
\end{align}
where $S^y_{tot}=\int d^2xS^y$, $s^y_{tot}=\int d^2xs^y$, and V is the two-dimensional volume of the interface.
The expected value of $j^{S^y}_z$ in the presence of the electric field $\bm{E}=(E_x,0,0)$ is given by the Kubo formula
\begin{align}
\langle j^{S^y}_z \rangle=\left[\lim_{\omega\rightarrow 0}\frac{K^y(\omega+i0)-K^y(0)}{i\omega}\right] E_x,\label{kuboformula}
\end{align}
where $K^y(\omega)$ is obtained from 
\begin{align}
K^y(i\omega_n)=\int_{0}^{1/T}d\tau e^{i\omega_n \tau}\langle \mathrm{T}_{\tau}j^{S^y}_z(\tau)j_x  \rangle
\end{align}
by the analytic continuation $i\omega_n\rightarrow\omega+i0$.
Here $j^x\equiv e\int d^2k/(2\pi)^2\psi^{\dagger}_{\bm{k}}\partial_{k_x}\hat{H}_e(\bm{k})\psi_{\bm{k}}$.

In the case of the conventional spin-orbit torque, lowest-order contributions to Eq. ($\ref{kuboformula}$) are $\mathcal{O}(J_{sd})$:
\begin{align}
\langle j^{S^y}_z \rangle\propto \epsilon_{yjk} \langle \psi^{\dagger}\hat{\sigma}_k\psi\rangle \langle S^j\rangle_{eq}+\mathcal{O}(J_{sd}^2),
\end{align}
where $\langle S^j\rangle_{eq}$ is the equilibrium expectation value of the FM spin.
In our case, on the other hand, $\mathcal{O}(J_{sd})$ contributions do not exist
since $\langle S^{x}\rangle_{eq}=\langle S^{z}\rangle_{eq}=0$.
The lowest-order ($J_{sd}^2$) contributions to $K^y(i\omega_n)$ are expressed diagrammatically in Fig. $\ref{fig2}$. 
(See Appendix \ref{dervofgreen} for details of the calculation.)
In this section, we drop impurity vertex corrections that are discussed in Sec. \ref{vertex}.
The contribution from Fig. $\ref{fig2}$(a), $K^{y(a)}(i\omega_n)$, is given by 
\begin{align}
K^{y(a)}(i\omega_n)=&ev\frac{J_{sd}^2a^4}{8}T^2\sum_{\nu,\omega_m}\left(\sqrt{\frac{2}{L}}\right)^2\sum_{q_n}\sum_{\alpha,\beta,\gamma=\pm}\notag\\
&\int\frac{d^2kd^2k'}{(2\pi)^2(2\pi)^2}\epsilon_{yjk}\chi^{lj}_{\bm{k'}-\bm{k},q_n}(i\omega_m)\notag\\
&\mathrm{Tr}\left[\hat{\sigma}_y|\bm{k},\alpha\rangle\langle \bm{k},\alpha|\hat{\sigma}_k|\bm{k'},\beta\rangle\langle \bm{k'},\beta|\hat{\sigma}_l|\bm{k},\gamma\rangle\langle \bm{k},\gamma|\right]\notag\\
&g_{\bm{k},\alpha}(i\nu)g_{\bm{k'},\beta}(i\nu+i\omega_m-i\omega_n)g_{\bm{k},\gamma}(i\nu-i\omega_n),\label{kernel}
\end{align} 
where $\chi^{lj}_{\bm{k'}-\bm{k},q_n}(i\omega_m)$ is the spin-spin correlation function.
In the third- and fourth- lines, we use Eq. ($\ref{electrongreen}$).
The third-line factor can be interpreted as the transition probability of electron-magnon scattering processes in the spin-momentum-locked bands. 
A similar expression is obtained for Fig. $\ref{fig2}$(b).

In the following, we focus on the scattering processes on the Fermi surface, which contribute dominantly to the diffusive phenomenon, and set $\alpha=\beta=\gamma=+$ [Fig. \ref{fig1}(c)].
By using standard analytic continuation techniques (Appendix \ref{appendixa}) and the relationship
\begin{align}
\chi^{xx}_{\bm{q},q_n}(i\omega_m)+\chi^{zz}_{\bm{q},q_n}(i\omega_m)=S_0\left[\mathcal{D}_{\bm{q},q_n}(i\omega_m)+\mathcal{D}_{\bm{q},q_n}(-i\omega_m)\right],
\end{align}
we obtain
\begin{widetext}
\begin{align}
\langle j^{S^y}_z \rangle_0=&-E_x\frac{\pi J_{sd}^2a^5S_0ev\tau}{8}\int\frac{d^2kd^2k'}{(2\pi)^2(2\pi)^2}\int\frac{2dq}{\pi}\frac{\partial f}{\partial \xi_{\bm{k}}}\cos\theta_{\bm{k}}(\cos\theta_{\bm{k}}-\cos\theta_{\bm{k'}})\notag\\
&\left[\left\{n_B(\omega_{\bm{k}-\bm{k'},q})+f(\xi^+_{\bm{k'}}) \right\}\delta(\xi^+_{\bm{k}}-\xi^+_{\bm{k'}}+\omega_{\bm{k}-\bm{k'},q})
+\left\{n_B(\omega_{\bm{k}-\bm{k'},q})+1-f(\xi^+_{\bm{k'}}) \right\}\delta(\xi^+_{\bm{k}}-\xi^+_{\bm{k'}}-\omega_{\bm{k}-\bm{k'},q})\right],\label{mainresult}
\end{align}
\end{widetext}
where $n_B(x)$ and $f(x)$ are the Bose and Fermi distribution functions, respectively.
$\langle\rangle_0$ denotes the expectation value without vertex corrections.
We replace $\sum_{q_n}$ with $(La/\pi)\int dq$ in the limit as $L\rightarrow\infty$.
The spin current discussed in this paper is mainly generated by the electron-magnon scattering between the opposite sides of the Fermi surface with the 
opposite spin directions.
The spin current generated by the electron-magnon interaction has been studied in Ref. [\onlinecite{mahfouzi}], which is only nonzero for a finite bias voltage due to the absence of the spin-momentum locking.

\section{Vertex corrections\label{vertex}}
\begin{figure}[t!]
\begin{center}
　　　\includegraphics[width=8cm,angle=0,clip]{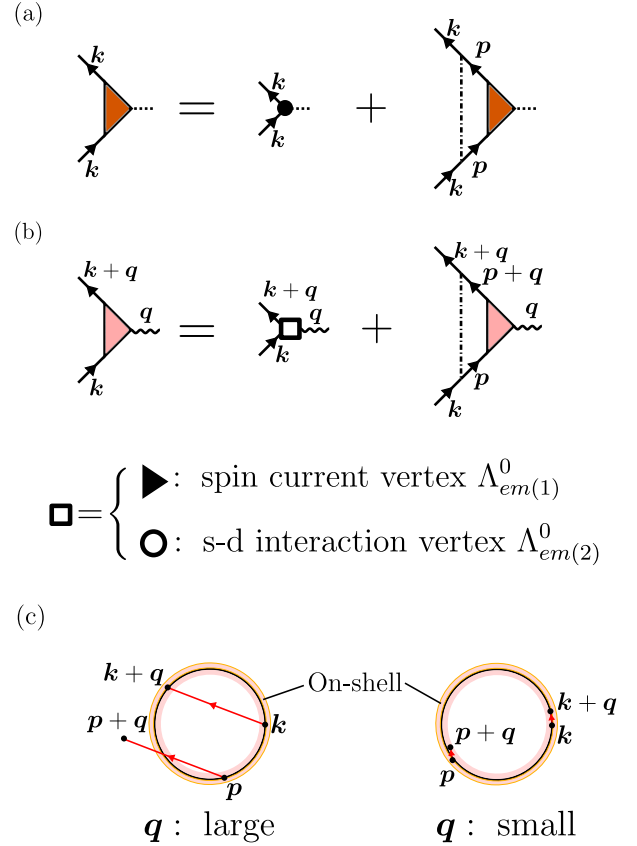}
　　　\caption{The impurity vertex corrections to (a) the electron-external field vertex and to (b) the electron-magnon vertex.
The chain lines denote impurity scatterings.
            (c) Schematic pictures of on-shell and off-shell scatterings.
            For sufficiently large $\bm{q}$, almost all scatterings $\bm{p}\rightarrow\bm{p}+\bm{q}$ ($|\bm{p}|\sim|\bm{k}|$) are off-shell.
            For sufficiently small $\bm{q}$, on the other hand, almost all scatterings are on-shell.
            }
　　　\label{fig3}
\end{center}
\end{figure}

In this section, we discuss the impurity vertex corrections in the Born approximation.
The corrections to the electron-external field vertex and the electron-magnon vertexes are shown in Figs. $\ref{fig3}$(a) and $\ref{fig3}$(b), respectively.

The self-consistent equation that corresponds to Fig. $\ref{fig3}$(a) is given by
\begin{align}
\hat{\Lambda}_{eE}(\bm{k})=\hat{\Lambda}^0_{eE}(\bm{k})+\frac{2}{\nu}\int &\frac{d^2p}{(2\pi)^2}\delta (\mu-v|\bm{p}|)\notag\\ 
&|\bm{p},+\rangle\langle \bm{p},+|\hat{\Lambda}_{eE}(\bm{p})|\bm{p},+\rangle\langle \bm{p},+|,\label{eEvertex}
\end{align}
where $\nu$ is the density of states at the Fermi energy, and $\hat{\Lambda}^0_{eE}=-v\sigma_y$ and $\hat{\Lambda}_{eE}$ are the bare and renormalized electron-external field vertexes, respectively. By solving Eq. ($\ref{eEvertex}$), we obtain\cite{adroguer,raghu,burkov}
\begin{align}
\hat{\Lambda}_{eE}(\bm{k})=2\hat{\Lambda}^0_{eE}(\bm{k}).
\end{align}

The self-consistent equation that corresponds to Fig. $\ref{fig3}$(b) is given by
\begin{align}
&\hat{\Lambda}_{em(i)}(\bm{k},\bm{k}+\bm{q})=\hat{\Lambda}^0_{em(i)}(\bm{k},\bm{k}+\bm{q})\notag\\
&+\frac{1}{\pi\nu\tau}\int \frac{d^2p}{(2\pi)^2}\frac{1}{(\mu-v|\bm{p}|+\frac{i}{2\tau})(\mu-v|\bm{p}+\bm{q}|-\frac{i}{2\tau})}\notag\\ 
&|\bm{p},+\rangle\langle \bm{p},+|\hat{\Lambda}_{em(i)}(\bm{p},\bm{p}+\bm{q})|\bm{p}+\bm{q},+\rangle\langle \bm{p}+\bm{q},+|,\label{emvertex}
\end{align}
where $\hat{\Lambda}^0_{em(i)}$ and $\hat{\Lambda}_{em(i)}$ are the bare and renormalized electron-magnon vertexes, respectively.
The explicit forms of $\hat{\Lambda}^0_{em(i)}$ are given by
\begin{align}
\hat{\Lambda}^0_{em(1)}(\bm{k},\bm{k}+\bm{q})=-\frac{J_{sd}a^2}{4}(S^z_{\bm{q}}\hat{\sigma}_x-S^x_{\bm{q}}\hat{\sigma}_z),\notag\\
\hat{\Lambda}^0_{em(2)}(\bm{k},\bm{k}+\bm{q})=-\frac{J_{sd}a^2}{2}(S^x_{\bm{q}}\hat{\sigma}_x+S^z_{\bm{q}}\hat{\sigma}_z),
\end{align}
where $i=1,2$ denote the spin current and the $s$-$d$ interaction vertexes, respectively.
It is not easy to compute these corrections since the vertexes depend on the magnon momentum $\bm{q}$.
Instead of the direct calculation, we here give a brief discussion.
In the limit $\mu\gg\omega_{\bm{q},q}$, which we consider in the next section, the electron-magnon scattering $\bm{k}\rightarrow\bm{k}+\bm{q}$ is approximately on-shell.
For sufficiently large $\bm{q}$, the amplitude of the scatterings $\bm{p}\rightarrow\bm{p}+\bm{q}$ ($|\bm{p}|\sim|\bm{k}|$) are negligible since the scatterings are off-shell [Fig. $\ref{fig3}$(c)].
Thus, the vertex corrections are suppressed due to the small value of the internal-momentum integral in Eq. ($\ref{emvertex}$).
The above discussion can not be applied to small $\bm{q}$.
For sufficiently small $\bm{q}$, almost all scatterings $\bm{p}\rightarrow\bm{p}+\bm{q}$ ($|\bm{p}|\sim|\bm{k}|$) are approximately on-shell [Fig. $\ref{fig3}$(c)].
In this case, the vertex corrections for small $\bm{q}$ are not negligible, since the self-consist equation has the same form as Eq. ($\ref{eEvertex}$):
\begin{align}
\hat{\Lambda}_{em(i)}(\bm{k},\bm{k}+\bm{q})\simeq&\hat{\Lambda}_{em(i)}(\bm{k},\bm{k})\notag\\
=&\hat{\Lambda}^0_{em(i)}(\bm{k},\bm{k})+\frac{2}{\nu}\int \frac{d^2p}{(2\pi)^2}\delta (\mu-v|\bm{p}|)\notag\\ 
&|\bm{p},+\rangle\langle \bm{p},+|\hat{\Lambda}_{em(i)}(\bm{p},\bm{p})|\bm{p},+\rangle\langle \bm{p},+|.\label{smallq}
\end{align}
By solving Eq. ($\ref{smallq}$), we obtain
\begin{align}
&\hat{\Lambda}_{em(1)}(\bm{k},\bm{k}+\bm{q})\simeq-\frac{J_{sd}a^2}{2}S^z_{\bm{q}}\hat{\sigma}_x+\frac{J_{sd}a^2}{4}S^x_{\bm{q}}\hat{\sigma}_z,\notag\\
&\hat{\Lambda}_{em(2)}(\bm{k},\bm{k}+\bm{q})\simeq -J_{sd}a^2S^x_{\bm{q}}\hat{\sigma}_x-\frac{J_{sd}a^2}{2}S^z_{\bm{q}}\hat{\sigma}_z.
\end{align}

Although there are the corrections for small $\bm{q}$, we here drop the electron-magnon vertex corrections.
To justify this approximation, we rewrite Eq. ($\ref{mainresult}$) as follows:
\begin{align}
\langle j^{S^y}_z\rangle_0=&\int\frac{d^2k}{(2\pi^2)}\cos\theta_{\bm{k}}\notag\\
&\left[\int_{\bm{q}:\mathrm{small}}+\int_{\bm{q}:\mathrm{large}}   \right](\cos\theta_{\bm{k}}-\cos\theta_{\bm{k}+\bm{q}})A^0(|\bm{k}|,\bm{q}),
\end{align}
where $A^0$ is a $\theta_{\bm{k}}$-independent function.
Since the contribution from $\int_{\bm{q}:small}$ is small due to the factor $(\cos\theta_{\bm{k}}-\cos\theta_{\bm{k}+\bm{q}})$,
the electron-magnon vertex corrections to $A^0$ for small $\bm{q}$ do not change the spin current expression drastically. 

In the following, we include only the correction to the electron-external field vertex and use the following expression for the spin current:
\begin{align}
\langle j^{S^y}_z \rangle=2\times\langle j^{S^y}_z\rangle_0.\label{cor}
\end{align} 

\section{Explicit calculation\label{explicit}}
To evaluate Eq. ($\ref{cor}$), we assume that $\mu\gg\omega_{\bm{q},q},T$.
In this limit, the following approximations are valid:
\begin{subequations}
\begin{align}
&\frac{\partial f}{\partial \xi^+_{\bm{k}}}\simeq-\delta(\xi^+_{\bm{k}}),\\
&\delta(\xi^+_{\bm{k}}-\xi^+_{\bm{k'}}\pm\omega_{\bm{k}-\bm{k'},q})\simeq \delta(\xi^+_{\bm{k}}-\xi^+_{\bm{k'}}),\\
&\omega_{\bm{k}-\bm{k'},q}\simeq Dq^2+2Dk_F^2[1-\cos(\theta_{\bm{k}}-\theta_{\bm{k'}})],
\end{align}\label{approx}
\end{subequations}
where $k_F=\mu/v$ is the Fermi wave number.

In the following, we evaluate Eq. ($\ref{cor}$) for the two limits: $T\gg\omega_{\bm{q},q}$ and $T\ll\omega_{\bm{q},q}$. (See Appendix \ref{appendixb} for details of the calculation.)
\subsection{High-temperature limit ($T\gg\omega_{\bm{q},q}$)}
Using Eqs. ($\ref{approx}$) and $n_B(\omega_{\bm{q},q})\simeq T/\omega_{\bm{q},q}\gg 1$,
we obtain the following expression for Eq. ($\ref{cor}$):
\begin{align} 
\langle j^{S^y}_z \rangle=\frac{J_{sd}^2a^5S_0(k_BT)\tau}{\hbar^2vD}\frac{ek_F}{8\pi^2}E_x,\label{spinhallcoeff}
\end{align}
where we insert $\hbar$ and $k_B$ for convenience.
Note that the ratio of the spin current to the electric current does not depend on the chemical potential:
\begin{align}
\frac{e\langle j^{S^y}_z \rangle}{\langle j_x\rangle}=\frac{J_{sd}^2a^5S_0(k_BT)}{2\pi\hbar v^2D},
\end{align}
where we use $\langle j_x\rangle=(e^2\mu\tau/4\pi\hbar) E_x$ \cite{dimension}.
\subsection{Low-temperature limit ($T\ll\omega_{\bm{q},q}$)}
In this limit, $f(\xi^+_{\bm{k'}})\simeq1-\Theta(|\bm{k'}|-k_F)\simeq1-\Theta(|\bm{k'}|-|\bm{k}|)$.
Thus, we obtain
\begin{align}
&f(\xi^+_{\bm{k'}})\delta(\xi^+_{\bm{k}}-\xi^+_{\bm{k'}}+\omega_{\bm{k}-\bm{k'},q})\notag\\
&+[1-f(\xi^+_{\bm{k'}})]\delta(\xi^+_{\bm{k}}-\xi^+_{\bm{k'}}-\omega_{\bm{k}-\bm{k'},q})\simeq0.\label{cancel}
\end{align}
Using $n_B(\omega_{\bm{q},q})\simeq e^{-\omega_{\bm{q},q}/T}$, Eqs. ($\ref{approx}$), and Eq. ($\ref{cancel}$), we obtain
\begin{align}
\langle j^{S^y}_z \rangle&=\frac{J_{sd}^2a^5S_0e\tau(k_BT)^2}{128\pi^2\hbar^2vk_FD^2}E_x,\\
\frac{e\langle j^{S^y}_z \rangle}{\langle j_x\rangle}&=\frac{J_{sd}^2a^5S_0(k_BT)^2}{32\pi \hbar v^2k_F^2D^2}.
\end{align}
In contrast to the high-temperature limit, the ratio depends on the chemical potential $\mu=vk_F$.
Also, the temperature dependence of the magnon spin current is different from that for high temperatures.

\section{Discussion and Summary\label{dissum}}

\begin{table}[]
\caption{List of material parameters and their typical values.}
\label{table1}
\begin{ruledtabular}
\begin{tabular}{lcc}
Quantity & Symbol& Value\\
\hline
Fermi wave number & $k_F$ &$10^9$ /m\\
Fermi velocity & $v$&$5\times 10^5$ m/s\\
Impurity relaxation time & $\tau$& $10^{-13}$ s\\
$s\mathchar`-d$ coupling & $J_{sd}$ & $10$ meV\\
Lattice constant & $a$ & $10^{-9}$ m\\
Stiffness &$D$&$5\times10^{-21}$ m$^2$ eV \\
Spin per unit cell&$S_0a^2$&$10$\\
Temperature&$k_BT$&30 meV
\end{tabular}
\end{ruledtabular}
\end{table}

As discussed in Sec. \ref{explicit}, the dominant terms to the spin current include the magnon distribution function $n_B(\omega_{\bm{q},q})$.
Thus, the spin current is enhanced at high temperatures where a lot of magnons contribute to the scattering processes.
For a quantitative estimate, we consider a TI/YIG heterostructure at room temperature.
Table \ref{table1} lists the parameters and their typical values.
We use the following typical values for Bi$_2$Se$_3$ and YIG: the Fermi velocity $v\sim5\times10^5$ m/s \cite{xlq,hzhang}, the impurity relaxation time $10^{-13}$ s \cite{taskin}, and the stiffness $D\sim 5\times10^{-21}$ m$^2$ eV\cite{shinozaki,srivastava}.
Experimentally, the Fermi wave number is a tunable parameter. We here choose $k_F\sim10^9$ /m. 
Although the value of the $s\mathchar`-d$ coupling $J_{sd}$ in this system is not well known, we adopt $J_{sd}\sim10$ meV obtained in the Supplemental Material of Ref. [\onlinecite{shiomi}].
For these parameters, $J_{sd}/\mu\sim10^{-1}$, which justifies our perturbation theory.
The spin quantum number of the spins in YIG, $S_0a^2$ in this paper, is $10$ per unit cell \cite{cherepanov}.
Using Eq. ($\ref{spinhallcoeff}$), we obtain $j^{S^y}_z/E_x\sim10^1$ ($\hbar/e$) $(\Omega\ \mathrm{cm})^{-1}$, whose dimension is the same as the spin Hall conductivity.
Although the phenomenon discussed in this paper differs from the spin Hall effect, it is interesting to note that this is an order of magnitude larger than the typical value for a semiconductor \cite{matsuzaka}.
Recently, spin current in a trilayer TI/Cu/FM was evaluated experimentally by means of the spin torque ferromagnetic resonance \cite{kondou}.
Except for the minor difference between the bilayer and the trilayer, our theory is expected to be experimentally accessible.

The same effect would occur in electron gas systems with the Rashba spin-orbit interaction due to the presence of two spin-momentum-locked 
Fermi surfaces. However, the large portion of the effect from the two Fermi surfaces with opposite chirality is reduced, and the spin-charge conversion efficiency would be lower than that of the TI/FM interface. 

In summary, we have studied electrical transport in a topological insulator/ferromagnet heterostructure in which the magnetization is perpendicular to the electric field. 
We have derived the expressions of the spin current induced by the coupling between the spin-momentum-locked surface state and magnons.
Using the parameters of Bi$_2$Se$_3$ and yttrium iron garnet, we have obtained $j^{S^y}_z/E_x\sim10^1$ ($\hbar/e$) $(\Omega\ \mathrm{cm})^{-1}$ at room temperature.
\begin{acknowledgments}
This work is supported by the World Premier International Research Center Initiative (WPI) and Grants-in-Aid for Scientific Research (No. JP15H05854, No. JP26400308, and No. JP25247056) from the Ministry of Education, Culture, Sports, Science and Technology (MEXT), Japan.
K. N. acknowledges many fruitful discussions with Yuki Shiomi, Eiji Saitoh, and Yoichi Ando.
N. O. acknowledges many fruitful discussions with Tomoki Hirosawa.
N. O. is supported by the Japan Society for the Promotion of Science (JSPS) through the Program for Leading Graduate Schools (MERIT).
N. O. is also supported by JSPS KAKENHI (Grants No. 16J07110).
\end{acknowledgments}
\appendix
\section{Calculation of $\mathcal{O}(J_{sd}^2)$ contributions to $K^y(i\omega_n)$ \label{dervofgreen}}
We here calculate the correlation function Eq. (\ref{kernel}).
To simplify the notation, we use $\sum_K$ for three-dimensional momentum summation.
In the imaginary time representation, the correlation function is calculated as follows:
\begin{widetext}
\begin{align}
\langle\mathrm{T}_{\tau}j^{S^y}_z(\tau)j_x\rangle&=ev\frac{J_{sd}}{4}\epsilon_{yjk}(\hat{\sigma}_k)_{ab}(\hat{\sigma}_y)_{cd}\sum_{K,K',Q}
\langle\mathrm{T}_{\tau}S^j_Q(\tau)\psi^{\dagger}_{K+Q,a}(\tau)\psi_{K,b}(\tau)\psi^{\dagger}_{K',c}(0)\psi_{K',d}(0)  \rangle\notag\\
&=\mathcal{O}(J_{sd})+ev\frac{J_{sd}^2a^4}{8}\epsilon_{yjk}(\hat{\sigma}_k)_{ab}(\hat{\sigma}_y)_{cd}(\hat{\sigma}_l)_{ef}\int_{0}^{1/T}d\tau'\sum_{K,K',K'',Q}\langle\mathrm{T}_{\tau}S^j_{Q}(\tau)S^l_{-Q}(\tau')\rangle_0\notag\\
&\ \ \ \ \ \ \ \ \ \ \ \ \ \ \ \langle\mathrm{T}_{\tau}\psi^{\dagger}_{K+Q,a}(\tau)\psi_{K,b}(\tau)\psi^{\dagger}_{K''-Q,e}(\tau')\psi_{K'',f}(\tau')\psi^{\dagger}_{K',c}(0)\psi_{K',d}(0)\rangle_0+\cdots.
\end{align}
\end{widetext}
Using the Wick's theorem and the Matsubara Fourier transformation, we obtain the $\mathcal{O}(J_{sd}^2)$ contributions [see Eq. (\ref{kernel}) and Fig. \ref{fig2}] to $K^y(i\omega_n)$ in terms of Green's functions Eqs. ($\ref{electrongreen}$) and ($\ref{magnongreen}$).

\section{Matsubara sum\label{appendixa}}
Here we perform the summation over the Matsubara frequencies in Fig. $\ref{fig2}$.
\subsection{Summation over the Bosonic Matsubara frequencies}
In the following, we calculate the summation $S_1=T\sum_{\omega_m}\left[\mathcal{D}_{\bm{q},q}(i\omega_m)-\mathcal{D}_{\bm{q},q}(-i\omega_m)\right]g_{\bm{k'}}(i\lambda+i\omega_m)$, where $\lambda=\nu-\omega_n$ for Fig. $\ref{fig2}$(a) and $\lambda=\nu$ for Fig. $\ref{fig2}$(b), respectively.
By using an analytic continuation technique, we obtain
\begin{widetext}
\begin{align}
S_1&=T\sum_{\omega_m}\frac{1}{i\omega_m-\omega_{\bm{q},q}} \frac{1}{i\lambda+i\omega_m-\xi^+_{\bm{k'}}}-T\sum_{\omega_m}\frac{1}{i\omega_m+\omega_{\bm{q},q}} \frac{1}{i\lambda+i\omega_m-\xi^+_{\bm{k'}}}\notag\\
&=\oint \frac{dz}{2\pi i}n_B(z)\frac{1}{z-\omega_{\bm{q},q}}\frac{1}{z+i\lambda-\xi^+_{\bm{k'}}}-\oint \frac{dz}{2\pi i}n_B(z)\frac{1}{z+\omega_{\bm{q},q}}\frac{1}{z+i\lambda-\xi^+_{\bm{k'}}}\notag\\
&=\frac{n_B(\omega_{\bm{q},q})-n_B(\xi^+_{\bm{k'}}-i\lambda)}{i\lambda-(\xi^+_{\bm{k'}}-\omega_{\bm{q},q})}
-\frac{n_B(-\omega_{\bm{q},q})-n_B(\xi^+_{\bm{k'}}-i\lambda)}{i\lambda-(\xi^+_{\bm{k'}}+\omega_{\bm{q},q})}\notag\\
&=\frac{n_B(\omega_{\bm{q},q})+f(\xi^+_{\bm{k'}})}{i\lambda-(\xi^+_{\bm{k'}}-\omega_{\bm{q},q})}
+\frac{n_B(\omega_{\bm{q},q})+1-f(\xi^+_{\bm{k'}})}{i\lambda-(\xi^+_{\bm{k'}}+\omega_{\bm{q},q})},
\end{align}
\end{widetext}
where we use $n_B(-x)=-n_B(x)-1$ and $n_B(x-i\lambda)=-f(x)$.
We omit $\mathrm{sgn}(\lambda)i/2\tau$.
\subsection{Summation over the Fermionic Matsubara frequencies}
\begin{figure}[]
\begin{center}
　　　\includegraphics[width=6cm,angle=0,clip]{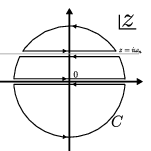}
　　　\caption{Path C in Eq. ($\mathrm{\ref{path}}$).}
　　　\label{fig4}
\end{center}
\end{figure}

In the following, we calculate the summation $S_2=T\sum_{\nu}g_{\bm{k}}(i\nu)g_{\bm{k}}(i\nu-i\omega_n)/(i\nu-i\omega_n-\xi''+\mathrm{sgn}(\nu-\omega_n)i/2\tau)$, where $\xi''=\xi^+_{\bm{k'}}\pm\omega_{\bm{q},q}$. 
In the presence of the impurity self-energy, the Matsubara sum can be calculated as follows:
\begin{widetext}
\begin{align}
S_2&=T\sum_{\nu}\frac{1}{i\nu-\xi+\mathrm{sgn}(\nu)i/2\tau}\frac{1}{i\nu-i\omega_n-\xi+\mathrm{sgn}(\nu-\omega_n)i/2\tau}\frac{1}{i\nu-i\omega_n-\xi''+\mathrm{sgn}(\nu-\omega_n)i/2\tau}\notag\\
&=-\oint_C\frac{dz}{2\pi i}\frac{1}{z-\xi+\mathrm{sgn}(\Re[z])i/2\tau}\frac{1}{i\nu-i\omega_n-\xi+\mathrm{sgn}(\Re[z]-\omega_n)i/2\tau}\frac{1}{i\nu-i\omega_n-\xi''+\mathrm{sgn}(\Re[z]-\omega_n)i/2\tau}\notag\\
&=-\int\frac{d\omega f(\omega)}{2\pi i}\left[G^{R}(\omega+i\omega_n,\xi)G^{R}(\omega,\xi)G^{R}(\omega,\xi'')-G^{R}(\omega+i\omega_n,\xi)G^{A}(\omega,\xi)G^{A}(\omega,\xi'')\right.\notag\\
&\ \ \ \ \ \ \ \ \ \ \ \ \ \ \ \ \ \ \ \ \ \left.+G^{R}(\omega,\xi)G^{A}(\omega-i\omega_n,\xi)G^{A}(\omega-i\omega_n,\xi'')-G^{A}(\omega,\xi)G^{A}(\omega-i\omega_n,\xi)G^{A}(\omega-i\omega_n,\xi'')\right].\label{path}
\end{align}
\end{widetext}
where $\Re[z]$ denotes the real part of the complex $z$, $\xi$ denotes $\xi^+_{\bm{k}}$, $C$ denotes the path described in Fig. $\ref{fig4}$, and $G^{R(A)}(\omega,x)=1/(\omega-x\pm i/2\tau)$ is the retarded (advanced) Green's function.
Keeping the $G^RG^AG^A$ terms, replacing $G^{R(A)}(\omega\pm i\omega_n,x)$ with $\pm\partial_{\omega}G^{R(A)}(\omega,x)$, and using $G^R(\omega,x)G^A(\omega,x)\simeq 2\pi\tau\delta(\omega-x)$, we obtain the expression in Eq. ($\ref{mainresult}$).
It is important to note that the imaginary parts of the Figs. $\ref{fig2}$(a) and Fig. $\ref{fig2}$(b) cancel each other out.
\section{Calculation of momentum integral\label{appendixb}}
Here we perform the momentum integration in Eq. ($\ref{mainresult}$).
Using $\theta\equiv\theta_{\bm{k}}$ and $\theta''\equiv\theta_{\bm{k}}-\theta_{\bm{k'}}$, we obtain
\begin{align}
\cos\theta_{\bm{k}}(\cos\theta_{\bm{k}}-\cos\theta_{\bm{k'}})=&\cos^2\theta(1-\cos\theta'')\notag\\
&-\sin\theta\cos\theta\sin\theta''.
\end{align}
Because the angular integration of the second-line term is zero, we keep the first-line term henceforth.
In the following, we replace $\int d\theta_{\bm{k}}d\theta_{\bm{k'}}$ with $\int d\theta d\theta''$.
\subsection{High-temperature limit}
As discussed in Sec. \ref{explicit}, we replace the two statistical factors, $\{n_B(\omega)+f(\xi^+)\}$ and $\{n_B(\omega)+1-f(\xi^+)\}$, with $T/\omega$.
Using Eqs. ($\ref{approx}$), we can perform the following integration:
\begin{widetext}
\begin{align}
&\int \frac{d\theta d\theta'' dkk dk'k'dq}{(2\pi)^2(2\pi)^2(\pi/2)}[-\delta(\xi^+_{\bm{k}})]\cos^2\theta(1-\cos\theta'')\frac{T}{Dq^2+2Dk_F^2(1-\cos\theta'')}2\delta(\xi^+_{\bm{k}}-\xi^+_{\bm{k'}})\notag\\
&=\frac{-T}{4\pi^4v^2D}\int d\theta'' dk k dk' k'dq(1-\cos\theta'')\frac{\delta(k-k_F)\delta(k'-k)}{q^2+2k_F^2(1-\cos\theta'')}\notag\\
&=\frac{-Tk_F^2}{4\pi^4v^2D}\int d\theta''(1-\cos\theta'')\frac{\pi}{2}\frac{1}{k_F\sqrt{2(1-\cos\theta'')}}\notag\\
&=\frac{-Tk_F}{2\pi^3v^2D}.
\end{align}
\end{widetext}
\subsection{Low-temperature limit}
As discussed in Sec. \ref{explicit}, we replace the two statistical factors, $\{n_B(\omega)+f(\xi^+)\}$ and $\{n_B(\omega)+1-f(\xi^+)\}$, with $e^{-\omega/T}$.
Using Eqs. ($\ref{approx}$), we can perform the integration as:
\begin{widetext}
\begin{align}
&\int \frac{d\theta d\theta'' dkk dk'k'dq}{(2\pi)^2(2\pi)^2(\pi/2)}(-\delta(\xi^+_{\bm{k}}))\cos^2\theta(1-\cos\theta'')\exp\left[-\frac{Dq^2+2Dk_F^2(1-\cos\theta'')}{T}\right]2\delta(\xi^+_{\bm{k}}-\xi^+_{\bm{k'}})\notag\\
&=\frac{-1}{4\pi^4v^2}\int d\theta'' dk k dk' k'dq(1-\cos\theta'')\exp\left[-\frac{Dq^2+2Dk_F^2(1-\cos\theta'')}{T}\right]\delta(k-k_F)\delta(k'-k)\notag\\
&=\frac{-k_F^2}{4\pi^4v^2}\int d\theta'' (1-\cos\theta'')\exp\left[-\frac{2Dk_F^2(1-\cos\theta'')}{T}\right]\int dq\exp\left[-\frac{Dq^2}{T}\right]\notag\\
&=\frac{-k_F^2}{4\sqrt{\pi^5}v^2}\sqrt{\frac{T}{D}}\exp\left[-\frac{2Dk_F^2}{T}\right]\left[I_0\left(\frac{2Dk_F^2}{T} \right)-I_1\left(\frac{2Dk_F^2}{T} \right)   \right]\notag\\
&\sim\frac{-k_F^3}{8\pi^3v^2}\left(\frac{T}{2Dk_F^2}\right)^2,
\end{align}
\end{widetext}
where $I_n(x)$ is the modified Bessel function of the first kind.
In the last line, we use the asymptotic form for $x\rightarrow \infty$: $[I_0(x)-I_1(x)]\sim e^x/(2x\sqrt{2\pi x})$.


\begin{thebibliography}{9}
\bibitem{hasan} M. Z. Hasan and C. L. Kane, Rev. Mod. Phys. $\bm{82}$, 3045 (2010).
\bibitem{xlq}X.-L. Qi and S.-C. Zhang, Rev. Mod. Phys. $\bm{83}$, 1057 (2011).
\bibitem{edelstein} V.M. Edelstein, Solid State Commun. $\bm{73}$, 233 (1990).
\bibitem{inoue}J. I. Inoue, G. E. W. Bauer, and L. W. Molenkamp, Phys. Rev. B $\bm{67}$, 033104 (2003).
\bibitem{kato}Y. K. Kato, R. C. Myers, A. C. Gossard, and D. D. Awschalom, Phys. Rev. Lett. $\bm{93}$, 176601 (2004).
\bibitem{silov}A. Yu. Silov,  P. A.  Blajnov, J. H. Wolter, R. Hey, K. H. Ploog, and N. S. Averkiev, Appl. Phys. Lett. $\bm{85}$, 5929 (2004).

\bibitem{nomura}K. Nomura and N. Nagaosa, Phys. Rev. Lett. $\bm{106}$, 166802 (2011).
\bibitem{sakai}A. Sakai and H. Kohno, Phys. Rev. B $\bm{89}$, 165307 (2014).
\bibitem{taguchi}K. Taguchi, K. Shintani, and Y. Tanaka, Phys. Rev. B $\bm{92}$, 035425 (2015).
\bibitem{yokoyama}T. Yokoyama, J. Zang, and N. Nagaosa, Phys. Rev. B $\bm{81}$, 241410(R) (2010). 
\bibitem{mahfouzi2}F. Mahfouzi, B. K. Nikoli\'{c}, and N. Kioussis , Phys. Rev. B $\bm{93}$, 115419 (2016).

\bibitem{fan}Y. Fan, P. Upadhyaya, X. Kou, M. Lang, S. Takei, Z.  Wang,  J.  Tang,  L.  He,  L.-T.  Chang,  M.  Montazeri, G. Yu, W. Jiang, T. Nie, R. N. Schwartz, Y. Tserkovnyak, and K. L. Wang, Nature Mater. $\bm{13}$, 699 (2014).
\bibitem{mellnik}A. R. Mellnik, J. S. Lee, A. Richardella, J. L. Grab, P. J. Mintun, M. H. Fischer, A. Vaezi, A. Manchon, E.-A. Kim, N. Samarth, and D. C. Ralph, Nature $\bm{511}$, 449 (2014).
\bibitem{ywang}Y. Wang, P. Deorani, K. Banerjee, N. Koirala, M. Brahlek, S. Oh,  and H. Yang, Phys. Rev. Lett. $\bm{114}$, 257202 (2015).


\bibitem{shiomi}Y. Shiomi, K. Nomura, Y. Kajiwara, K. Eto, M. Novak, K. Segawa, Y. Ando, and E. Saitoh, Phys. Rev. Lett. $\bm{113}$, 196601(2014).
\bibitem{hwang}H. Wang, J. Kally, J. S. Lee, T. Liu, H. Chang, D. R. Hickey, K. A. Mkhoyan, M. Wu, A. Richardella, and N. Samarth, Phys. Rev. Lett. $\bm{117}$, 076601 (2016).
\bibitem{yasuda} K. Yasuda, A. Tsukazaki, R. Yoshimi, K. S. Takahashi, M. Kawasaki, and Y. Tokura, Phys. Rev. Lett. $\bm{117}$, 127202 (2016).

\bibitem{kondou}K. Kondou, R. Yoshimi, A. Tsukazaki, Y. Fukuma, J. Matsuno, K. S. Takahashi, M. Kawasaki, Y. Tokura, and Y. Otani, Nat. Phys. $\bm{12}$, 1027 (2016).
\bibitem{dankert}A. Dankert, J. Geurs, M. V. Kamalakar, S. Charpentier and S. P. Dash, Nano Lett. $\bm{15}$, 7976 (2015).

\bibitem{tian}J. Tian, I. Miotkowski, S. Hong,  and Y. P. Chen, Scientific Reports $\bm{5}$, 14293 (2015).
\bibitem{jamali} M.  Jamali,  J.  S.  Lee,  J.  S.  Jeong,  F.  Mahfouzi,  Y.  Lv, Z. Zhao, B. K. Nikoli\'{c}, K. A. Mkhoyan, N. Samarth, and J.-P.  Wang, Nano Lett. $\bm{15}$, 7126 (2015).

\bibitem{jiang}Z. Jiang, C. Z. Chang, M. Ramezani Masir, C. Tang, Y. Xu, J. S. Moodera, A. H. MacDonald, and J. Shi, Nat. Commun. $\bm{7}$, 11458 (2016).

%\bibitem{fischer}M. H. Fischer, A. Vaezi, A. Manchon, and E. -A. Kim, Phys. Rev. B $\bm{93}$, 125303 (2016).
%\bibitem{ndiaye}P. B. Ndiaye, C. A. Akosa, M. H. Fischer, A. Vaezi, E-A. Kim and A. Manchon, arXiv:1509.06929. 


\bibitem{chernyshov} A. Chernyshov and M. Overby and X. Y. Liu and J. K. Furdyna,Y. Lyanda-Geller, and L. P. Rokhinson, Nature Phys. $\bm{5}$, 656 (2009).
\bibitem{miron}I. M. Miron, G. Gaudin, S. Auffret, B. Rodmacq, A. Schuhl, S. Pizzini, J. Vogel and P. Gambardella, Nature Mater. $\bm{9}$, 230 (2010).



\bibitem{takei} S.  Takei  and  Y.  Tserkovnyak,  Phys.  Rev.  Lett. $\bm{112}$, 227201 (2014).
\bibitem{takahashi}S. Takahashi, E. Saitoh, and S. Maekawa, J. Phys. Conf. Ser. $\bm{200}$, 062030 (2010).

\bibitem{adroguer}P. Adroguer, D. Carpentier, J. Cayssol,  and E. Orignac, New Journal of Physics $\bm{14}$, 103027 (2012).
\bibitem{raghu}S. Raghu, S.B. Chung, X.-L. Qi, and S.-C. Zhang, Phys. Rev. Lett. $\bm{104}$, 116401 (2010).
\bibitem{burkov}A. A. Burkov, D. G. Hawthorn, Phys. Rev. Lett. $\bm{105}$, 066802 (2010).

\bibitem{mahfouzi}
F. Mahfouzi and B. K. Nikoli\'{c}, Phys. Rev. B $\bm{90}$, 045115 (2014).

\bibitem{dimension}
Because the charge current density is defined on the surface state,
the dimension of it is [A/m], while the dimension of the spin current ($\times e$) is [A/m$^2$].


\bibitem{hzhang}H. Zhang, C.-X. Liu, X.-L. Qi, X. Dai, Z. Fang, and S.-C. Zhang, Nature Phys. $\bm{5}$, 438 (2009).
\bibitem{taskin}A. A. Taskin, S. Sasaki, K. Segawa, and Y. Ando, Phys. Rev. Lett. $\bm{109}$, 066803 (2012).
\bibitem{shinozaki}S. S. Shinozaki, Phys. Rev. $\bm{122}$, 388 (1961).
\bibitem{srivastava}C. M. Srivastava and R. Aiyar, J. Phys. C $\bm{20}$, 1119 (1987).
\bibitem{cherepanov}V. Cherepanov, I. Kolokolov,   and V. L'Vov, Physics Reports $\bm{229}$, 81 (1993).
\bibitem{matsuzaka}S. Matsuzaka, Y. Ohno, and H. Ohno, Phys. Rev. B $\bm{80}$, 241305(R) (2009).



\end{thebibliography}
\end{document}